\documentclass[lettersize,journal]{IEEEtran}
\usepackage{amsmath,amsfonts}
\usepackage{algorithmic}
\usepackage{algorithm}
\usepackage{array}
\usepackage[caption=false,font=normalsize,labelfont=sf,textfont=sf]{subfig}
\usepackage{textcomp,gensymb}
\usepackage{stfloats}
\usepackage{url}
\usepackage{verbatim}
\usepackage{graphicx}
\usepackage{cite}
\usepackage[table]{xcolor}

\usepackage{amssymb}
\usepackage{algorithmic}
\usepackage{enumitem}
\usepackage{xcolor}
\usepackage{soul}
\usepackage{booktabs}
\usepackage{hhline}
\usepackage{multirow}
\usepackage{boldline} 
\usepackage{array}

\newcolumntype{P}[1]{>{\centering\arraybackslash}p{#1}}
\newcolumntype{M}[1]{>{\arraybackslash}m{#1}}

\begin{document}

\title{Prospects and Applications of Incoherent Light\\ in Non-contact Wireless Sensing Systems}

\author{Md Zobaer Islam$^*$, Sabit Ekin$^*$,~\IEEEmembership{Senior Member, IEEE}, and John F. O'Hara, \IEEEmembership{Senior Member, IEEE}

\thanks{\textit{$^*$Corresponding authors: Md Zobaer Islam, Sabit Ekin}

Md Zobaer Islam and John F. O'Hara are with the School of Electrical and Computer Engineering, Oklahoma State University, Oklahoma, USA (e-mail: zobaer.islam, oharaj\{@okstate.edu\}).

Sabit Ekin is with the Department of Engineering Technology and Industrial Distribution, Texas A\&M University, Texas, USA (e-mail: sabitekin@tamu.edu).}

\thanks{This work has been submitted to the IEEE for possible publication. Copyright may be transferred without notice, after which this version may no longer be accessible.}

}

\markboth{IEEE Circuits and Systems Magazine,~Vol.~XX, No.~X, XXXX~2023}%
{Islam \MakeLowercase{\textit{et al.}}: Prospects and Applications of Incoherent Light in Wireless Sensing}


\maketitle

\begin{abstract}
The increasing demand for wireless sensing systems has led to the exploration of alternative technologies to overcome the spectrum scarcity of traditional approaches based on radio frequency (RF) waves or microwaves. Incoherent light sources such as light-emitting diodes (LED), paired with light sensors, have the potential to become an attractive option for wireless sensing due to their energy efficiency, longer lifespan, and lower cost. Although coherent light or laser may present safety risks to human eyes and skin, incoherent visible and infrared light has low intensity, and does not harm the human body. Incoherent light has the potential to supersede other wireless sensing technologies, namely RF, laser and camera, by providing many additional benefits including easy implementation, wide bandwidth, reusable frequency, minimum interference, enhanced privacy and simpler data processing. However, the application of incoherent light in the wireless sensing domain is still in its infancy and is an emerging research topic. This study explores the enormous potential and benefits of incoherent visible and infrared light in wireless sensing through various indoor and outdoor applications including speed estimation of vehicles, human vitals monitoring, blood glucose sensing, gesture recognition, occupancy estimation and structural health monitoring. 
\end{abstract}


\section{Introduction}
\IEEEPARstart{W}{ireless} sensing refers to the use of wireless technologies to detect and measure physical parameters of a target object or environment without having any physical contact between the sensor and the target. Examples of the physical parameters that can be measured may include temperature, pressure, humidity, motion, proximity, and other subtle movements present in the target. Wireless sensing is commonly utilized in applications where physical contact with the target is impractical or may alter the parameters to be measured. Wireless sensing has become an integral part of human life due to its ability to monitor and improve human health, enhance safety, and increase efficiency in various sectors. For example, wireless sensing technologies have revolutionized healthcare by enabling remote patient monitoring, early detection of health issues, and personalized treatments~\cite{shah2019rf, kranjec2014non, Jagadev2020}. In transportation, wireless sensing can optimize traffic flow, reduce congestion, and enhance vehicle safety~\cite{guerrero2018sensor}. In agriculture, wireless sensing can enhance crop yields, reduce resource waste, and improve sustainability~\cite{palazzi2019feeding, xu2020estimation}. Wireless sensing has also found extensive applications in many other sectors including industrial automation, security and surveillance, smart home and environmental monitoring.

\begin{figure}[b]
 \centering
 \includegraphics[width=0.5\textwidth]{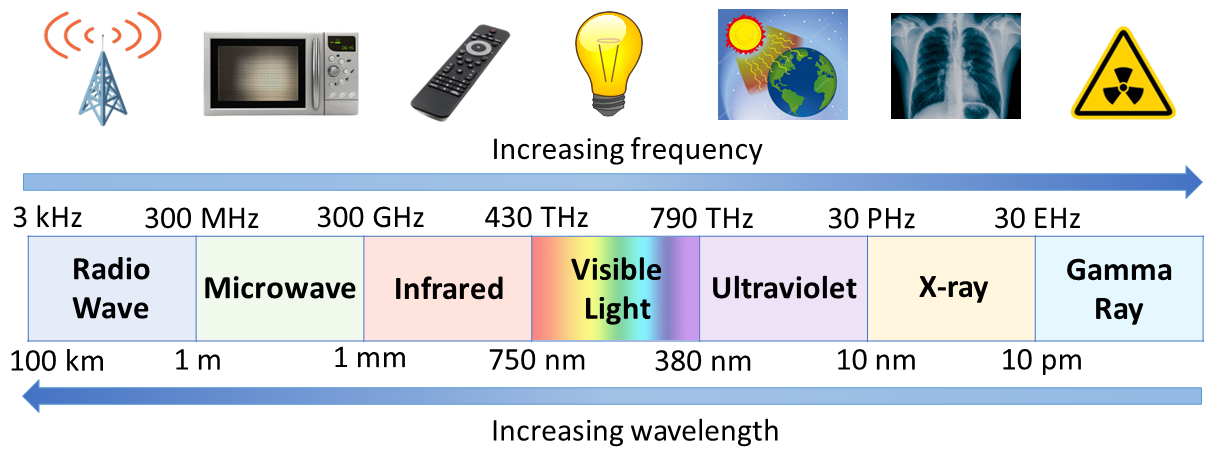}
 \caption{Electromagnetic spectrum.}
 \label{fig:emspectrum}
 \end{figure}

The conventional wireless sensing systems based on radio waves or microwaves~\cite{lubna2022radio, shah2019rf, ma2019wifi, basalamah2016, li2009radar, zhang2010, chen1986x}, employing frequencies ranging from 3\,kHz to 300\,GHz in the electromagnetic spectrum (Fig.~\ref{fig:emspectrum}), have encountered spectrum scarcity due to the growing number of high-bandwidth applications and their licensed users. Consequently, researchers have started to investigate alternative technologies, including optical frequencies (primarily laser)~\cite{kaartinen2022lidar, dervish2019position, perrin2004gesture, optglu3, zhang2020vehicle, kondo1997laser}, RGB or thermal infrared cameras~\cite{wilson2023recent, BarbosaPereira2017, tarassenko2014non, kumar2015distanceppg, moeslund2006survey, masoud2003method}, and sound waves~\cite{bai2020acoustic, qifan2014dolphin, watanabe2013ultrasound, na2015acoustic} for wireless sensing applications. In optical frequency-based wireless sensing and communication systems, unguided coherent or incoherent optical signals are transmitted through the atmosphere or free space and received by light sensors like photodiodes. It is worth noting that there is fundamental difference between coherent and incoherent light. Coherent light, produced by laser diodes, comprises light waves with fixed frequency and phase that propagate in unison toward a single direction. In contrast, incoherent light consists of light waves with random fluctuations in amplitude, frequency, phase, and polarization, and does not possess a fixed direction of propagation. Examples of incoherent light sources include the sun and other natural sources of light, incandescent bulbs, fluorescent lamps and most types of LEDs. While coherent light can pose safety risks to human eyes and skin~\cite{youker2001practical, parker2007laser, klingbeil1986safety}, incoherent light is ubiquitous, low-intensity, and usually poses no harm to the human body.

Incoherent light, including visible and infrared light, covers frequencies between 300\,GHz to 790\,THz (1000\,$\mu$m to 380\,$\mu$m in wavelength) on the electromagnetic spectrum (Fig.~\ref{fig:emspectrum}). Modern LED technology has made incoherent light an attractive option for wireless sensing and communication due to its higher energy efficiency, longer lifespan, and lower cost. In such applications, data is transmitted by modulating the light intensity of an optical source such as an LED. Infrared light is more suitable for discrete applications or where illumination is redundant, as it is invisible to human eyes. Visible light has already been used in a variety of wireless communication applications such as indoor, vehicular, power line, underground and underwater communications \cite{Matheus2019, Karunatilaka2015, Pathak2015}. There are ample opportunities to exploit incoherent light for wireless sensing too. A few studies used visible light for gesture, hand pose and human posture recognition and indoor localization, but they relied on shadow-based passive sensing approaches which required arrays of light sources and/or sensors~\cite{venkatnarayan2018gesture, li2018self, li2017reconstructing, di2016localight}. In some visible light-based indoor positioning applications, a smartphone camera was used as light sensor which could raise privacy concerns due to the potential for capturing unsolicited images~\cite{zhang2016litell, kuo2014luxapose}. Relevant studies on direct sensing of incoherent visible or infrared light without compromising privacy are limited so far in the literature. This study aims to discuss various aspects of using incoherent light for wireless sensing, along with its relevant applications in healthcare, transportation, building automation, human-computer interaction (HCI) and infrastructure management sectors. 

The remainder of this manuscript is organized as follows. Section~\ref{sec:advantages_of_VILS} describes the potential advantages of incoherent light-wave sensing (LWS). Section~\ref{sec:LWS_sys_model} presents the generic system model and some theory of LWS. Various applications of LWS are reviewed in detail in Section~\ref{sec:LWS_applications}, followed by some limitations and challenges of LWS in Section~\ref{sec:limitations_of_VILS}. Finally, Section~\ref{sec:conclusion} draws the conclusion and forecasts future research directions.

\begin{table*}[!htbp]
\renewcommand{\arraystretch}{1.3}
\centering
\caption{Comparison of incoherent light-based sensing with other related technologies.}
\begin{center}
\begin{tabular}{|M{3.1cm}|M{2.15cm} M{2.15cm} M{2.15cm}M{2.25cm}|} 
\hlineB{2}
\bf{Index} & \bf{RF} &  \bf{Coherent light}&  \bf{Optical camera} & \textbf{Incoherent light}\\
 \hhline{|=====|}
  {Safety} & Can be unsafe & Can be unsafe & \cellcolor{blue!17}Safer & \cellcolor{blue!17}Safer\\
  \hline
  {Spectrum licensing} & Required & \cellcolor{blue!17}Not required & \cellcolor{blue!17}Not required & \cellcolor{blue!17}Not required\\
  \hline
  {Privacy and security} & Limited & \cellcolor{blue!17}High & Limited & \cellcolor{blue!17}High\\
  \hline
  {System complexity} & High & High & Moderate & \cellcolor{blue!17}Low\\
  \hline
   {Computational complexity} & High & \cellcolor{blue!17}Low & High & \cellcolor{blue!17}Low\\
    \hline
   {Divergence angle} & Narrow & Narrow & \cellcolor{blue!17}Wide & \cellcolor{blue!17}Wide\\
   \hline
   {EM interference} &High & \cellcolor{blue!17}Low & Not applicable & \cellcolor{blue!17}Low\\
   \hline
   {Interference with light} &\cellcolor{blue!17}Low & High & Not applicable & High\\
   \hline
   {Precise alignment} &Required &Often required &\cellcolor{blue!17}Not required & \cellcolor{blue!17}Not required\\
   \hline
{Modulation bandwidth} &\cellcolor{blue!17}High &\cellcolor{blue!17}High &Not applicable & Low\\
\hline
   {Spectral width} &\cellcolor{blue!17}Narrow &\cellcolor{blue!17}Narrow &Not applicable & Wide\\
   \hline
   {Multipath fading} &High &\cellcolor{blue!17}Low &Not applicable & \cellcolor{blue!17}Low\\
   \hline
   {Shadowing} &\cellcolor{blue!17}Moderate &High &Not applicable & High\\
   \hline
   {Sensing range} &\cellcolor{blue!17}Long &\cellcolor{blue!17}Long &Medium & Short\\
\hline
   {Implementation cost} &High &High &Moderate & \cellcolor{blue!17}Low\\
\hlineB{2}
\end{tabular}
\end{center}
\label{table:comparison}
\end{table*}

\section{Advantages of Incoherent LWS}
\label{sec:advantages_of_VILS}
There are many advantages of incoherent light-wave sensing systems over other existing technologies like RF, laser or optical camera-based systems which are discussed next:
\begin{enumerate}[leftmargin=*]
\item \textbf{Wide bandwidth}: RF and microwave frequencies cover 3\,kHz to 300\,GHz frequencies in the electromagnetic spectrum. However, the bandwidth offered by visible and infrared light spectra is 2600 times greater than the combined bandwidth of RF and microwave signals.~\cite{haas2020}. As a result, visible light and IR bands can address the RF spectrum crisis.

\item \textbf{Non-licensed channels}: Obtaining a license from the Federal Communications Commission (FCC) is often an overhead in RF sensing and communication, which also restricts the reuse of frequencies by multiple applications. In contrast, frequencies of light waves are not licensed and are free to be used by anyone ~\cite{Jovicic2013}.


\item \textbf{Wide divergence angle}: LEDs and light sensors can have wide beamwidth and field of view, respectively, enabling diffused mode communication with poor directionality~\cite{Jungnickel2002}. 
In contrast, RF and coherent light-based sensing and communication often require precise alignment between transmitter and receiver.

\begin{figure*}[!htbp]
 \centering
  \includegraphics[width=0.98\textwidth]{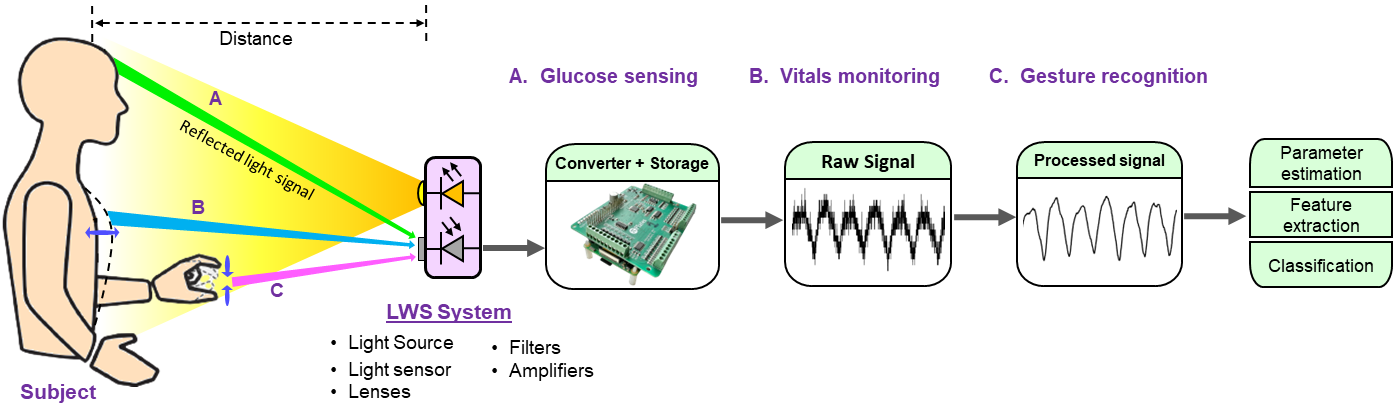}
 \caption{LWS system model for human-light interaction.}
 \label{fig:sysmodel_human}
 \end{figure*}

\item \textbf{Minimum electromagnetic interference}: Visible and infrared lights produce minimal RF radiation, posing no threat of electromagnetic interference. Thus, the technology is suitable for use in hospitals, airplanes and industrial environments where multiple RF devices are already present~\cite{Karunatilaka2015, Abuella2021}.

\item \textbf{Privacy and security}: The shorter wavelengths of visible and infrared lights compared to radio frequency (RF) wavelengths make them unable to penetrate opaque obstacles like walls. Thus, light can only be received in a confined space which ensures enhanced security and confidentiality of transmitted information~\cite{Karunatilaka2015,Pathak2015,classen2016, Kahn1997}. In optical camera-based sensing approaches, photos or videos of the subjects are regularly captured using red-green-blue (RGB) or thermal infrared camera which is a privacy breach, if it is done without prior permission. But in visible and infrared light sensing (VILS), the light sensor is a single-pixel light detector that is unable to capture two or higher dimensional images or videos, thereby ruling out privacy concerns.

\item \textbf{Health safety}: The use of RF signals and lasers can pose a health safety risk when the power level exceeds the allowable limit. The FCC regulates the power level for each RF device individually, but the combined power of multiple RF devices in close proximity can still be harmful to human health. Lasers used in optical wireless communication require strict power regulation to ensure the safety of human eyes and skin ~\cite{soltani2022,Kahn1997}. On the contrary, incoherent visible and infrared light signals are safe for the human body under normal lighting conditions. 

\item \textbf{Easy implementation}: Both visible and infrared lights are readily available in our environment and can be directly detected using light sensors. They are less affected by multipath fading and Doppler shift~\cite{Pathak2015,Jovicic2013, Barry1995, Kahn1997}, resulting in transmitters and receivers (sensors) of incoherent light being less complex and cheaper than their RF counterparts. In some cases, existing lighting infrastructure used for illumination can be utilized for sensing and communication, further reducing implementation overhead and propagation path loss.

\item \textbf{Simple data processing}: The data generated by VILS systems consist of 1-dimensional time-series data, which require simpler processing compared to the 2-dimensional time-frequency data obtained with RADARs or captured videos/images with optical camera-based sensing systems.
\end{enumerate}

Table~\ref{table:comparison} summarizes a comparison of various technologies from a wireless sensing perspective, with shaded cells indicating a comparative advantage over other technologies. Although the column on incoherent light-based sensing has the maximum number of shaded cells, it also has a few limitations, which are discussed in Section~\ref{sec:limitations_of_VILS} later on.

\section{The LWS System Model}
\label{sec:LWS_sys_model}
The proposed sensing system based on light waves comprises of an LED as the light source for transmitting incoherent visible or infrared light towards a target, and a light sensor or photodetector for receiving the reflected light signal. In case the target already has a light source, it can be used as a transmitter for sensing purposes. Additionally, the existing LED lights in the surroundings can also be utilized as transmitters depending on the specific application. The light sensor functions as a square-law detector, converting the received light intensity into voltage amplitudes, which are then digitally stored as time-series data. For the proposed light-Wave sensing system, p-i-n photodiodes are preferred over avalanche photodiodes due to their better quantum efficiency and reduced sensitivity towards shot noise induced by ambient light~\cite{Carruthers2002}.

Light propagation from visible or infrared LEDs follows the Lambertian propagation model~\cite{gfeller1979,Pathak2015}. According to this model, if $P_t$ is the optical power transmitted from a point source, then the power at a distance $D$, is given by
\begin{align}
   \label{eq:Lambertian_Channel1}
    P_D=\frac{(n+1)A P_t}{2\pi D^\gamma}\cos^n(\phi)\cos(\theta) , ~~\forall\theta< \phi_{1/2},
\end{align}
where $A$ is the area intercepted at the photodetector, $\gamma$ is the empirical path-loss exponent, $\phi$ is the irradiance angle, $\theta$ is the incident angle~\cite{abuella2017symposium, Abuella2019, Abuella2020} and $\phi_{1/2}$ is the half-power angle in the field of view of the light source. $n$ is the order of the Lambertian model and is given by $n=-\ln(2)/\ln\{\cos(\phi_{1/2})\}$.

If $P_r$ is the optical power received by the photodetector (either directly or reflected from a target), $R_{pd}$ is the responsivity and $i_d$ is the dark current of the photodetector\footnote{The responsivity of a light sensor is defined as the photocurrent generated per unit optical incident power on it. Dark current is the small amount of direct current present due to the thermal excitation of electrons in the light sensor, even when no light is incident on it.}, then the generated photocurrent becomes $i_{pd}=i_d+R_{pd}P_r$~\cite{detector2}. The transimpedance amplifier in the photodetector can convert this current into voltage
\begin{align}
\label{eq:voltage}
V_{r}=g_{pd}\left(i_d+R_{pd}P_r\right),
\end{align}
where $g_{pd}$ is the transimpedance gain. The voltage $V_{r}$ is fed to the analog-to-digital converter and subsequently stored digitally in a microcontroller or Raspberry Pi for offline processing. The received light intensity is modulated by small movements of the target, resulting in voltage data that contains corresponding fluctuations. These fluctuations can be utilized to extract useful information about the target. However, the received light is susceptible to environmental noise, such as ambient light or sunlight, and sensor noise, such as thermal and shot noise. These noises can be attenuated through offline digital signal processing on the collected data. The proposed LWS system model is illustrated in Fig.~\ref{fig:sysmodel_human} with a human subject. The collected raw and processed data can be used to estimate necessary parameters, extract features, and perform classification, pattern recognition, anomaly detection, or clustering through machine learning (ML) techniques. The subject can also be non-human, depending on the application.

\section{Applications}
\label{sec:LWS_applications}
Wireless visible light and infrared sensing offers numerous potential applications, such as vehicle-to-vehicle (V2V) and vehicle-to-everything (V2X) sensing and communication, indoor localization, occupancy estimation, structural health monitoring, and biomedical applications for monitoring human health parameters such as breathing rate, heart rate and blood glucose level. Many of these applications have already been investigated by researchers and will be reviewed in the following sub-sections.

\subsection{Vehicular Speed Estimation}
\label{subsec:vehicle_speed}
\begin{figure*}[!htbp]
 \centering
  \includegraphics[width=0.916\textwidth]{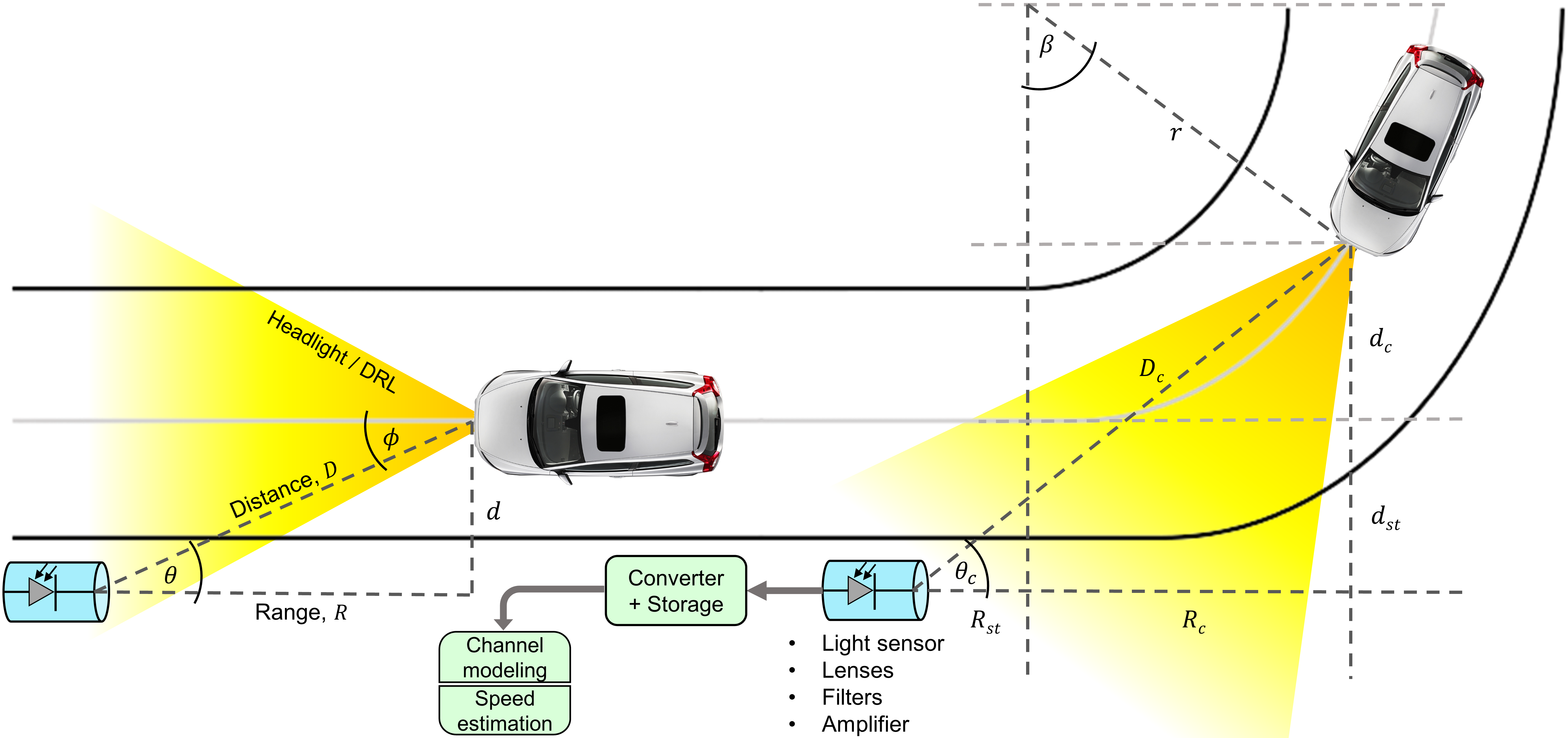}
 \caption{LWS system model for vehicular sensing in straight road and curved road scenarios.}
 \label{fig:sysmodel_vehicle}
 \end{figure*}

Wireless visible light and infrared sensing based on incoherent light have a variety of potential applications in the automotive and transportation industries. This is due to the fact that motor vehicles are already equipped with headlamps/tail-lamps and daytime running lights (DRLs) that can easily function as visible light transmitters. To ensure reliable and safe transportation, transportation and law enforcement agencies frequently need to estimate the speed of road traffic. Visible light sensing (VLS) using vehicle headlamps can provide a simpler solution with superior performance than existing RADAR/LiDAR-based speed estimation techniques. In~\cite{Abuella2019, abuella2019wireless, abuella2017symposium}, the authors proposed theoretical and algorithmic analyses and simulation results of vehicle speed estimation using VLS, which have also been patented~\cite{abuella2020Vildarpatent}. Fig.~\ref{fig:sysmodel_vehicle} depicts the proposed system model for vehicular sensing in both straight and curved road scenarios. The authors made the following reasonable assumptions in their analyses of visible light-based vehicular channel modeling and speed estimation of vehicles:
\begin{enumerate}
\item Only a single vehicle passes through the road (equivalent to focusing on a single lane).
\item The vehicle moves with constant velocity.
\item The vehicle's LED transmits a constant optical power.
\item The vehicle headlamp and roadside light sensor are at the same height. 
\item The light sensor is placed close to the road.
\item The distance between two headlamps of the vehicle is negligible i.e., they work like a single transmitter.
\end{enumerate}

\subsubsection{\textbf{Channel Modeling and Validation}}
\label{subsubsec:ch_modeling}
The light propagation channel can be modeled for vehicular sensing and communication using equation~(\ref{eq:Lambertian_Channel1}) with additional simplifying assumptions mentioned above~\cite{abuella2019wireless,Abuella2019}. When the vehicle headlight and light sensor are at the same height, and the distance between two headlights of the vehicle is negligible, the irradiance angle ($\phi$) and the incident angle ($\theta$) become equal in equation~(\ref{eq:Lambertian_Channel1}). Moreover, these angles become negligible when the detector is close to the road and the car is far away. Thus, we can have $\cos(\phi)=\cos(\theta)\approx1$. When the headlight is used as the light transmitter in the LWS system, the optical power at a distance $D$ is the power received by the light sensor, hence $P_D=P_r$. Then, equation~(\ref{eq:Lambertian_Channel1}) becomes
\begin{align}
   \label{eq:commonchannel}
    P_r=KD^{-\gamma},
\end{align}
where $K=\frac{(n+1)A P_t}{2\pi}$, a constant. This model resembles the typical RF path loss model shown in~\cite{rappaport2015}. $K$ and $\gamma$ can be estimated through simulation~\cite{Abuella2019} or experiment~\cite{abuella2023}.

In~\cite{abuella2023}, a simplified channel model for vehicular sensing and communication is proposed based on the Lambertian propagation model which is

\begin{equation}
 \label{eqn:conditiona_eq}
P_{r_{dB}}=\begin{cases}K_{dB} - \gamma D_{dB},&\frac{d^2}{D^2} \ll 1
\\K_{dB} - \gamma D_{dB} + G_{dB},&\text{otherwise} \end{cases}
 \end{equation}
where $P_{r_{dB}} = 10 \log_{10}(P_{r})$, $K_{dB} = 10 \log_{10}(K)$, $D_{dB} = 10 \log_{10}(D)$ and $G_{dB}= 5(n+1) \log_{10}\left(1-\frac{d^2}{D^2}\right)$.

The results of this study demonstrate a linear relationship between received power and distance in dB domain, with a slope of $\gamma$ observed in the far scenario (where $D\gg d$). This relationship is consistent with equation~(\ref{eq:commonchannel}) in dB domain. However, non-linearity is introduced in the near scenario by the term $G_{dB}$. To gather data for static channel modeling, outdoor measurement campaigns were conducted at different fixed distances between a vehicle and a roadside light sensor. Data were also collected during dynamic channel modeling as the vehicle moved continuously toward the light sensor. Analysis of received power (in dBW) vs. distance (in dB) plots under both night and sunny daylight conditions revealed that the data followed equation~(\ref{eqn:conditiona_eq}), thereby validating the proposed channel model experimentally~\cite{abuella2023}. 

\subsubsection{\textbf{Speed Estimation}}
\label{subsubsec:speed_est}
If the path loss model and the received power values with corresponding time instances are known, then equation~(\ref{eq:commonchannel}) can be used to find the distance values ($D$). The knowledge of distance-difference ($\Delta D$) and time-difference ($\Delta t$) between two measurements can yield instantaneous speed $V_i=\Delta D_i/\Delta t_i$, which is the basic approach of speed estimation using VLS. But this approach is highly dependent on individual received signal data affected by noise, hence it may yield a high percentage of error in speed estimation. It has been shown in~\cite{abuella2017symposium,Abuella2019} that under the assumption of constant speed of the vehicle during measurement time in the straight road scenario, the channel model can be rearranged to
\begin{align}
   \label{eq:lineareqn}
    \sqrt{\left(\frac{P_r}{K}\right)^{-\frac{2}{\gamma}}-d^2}=-Vt+R_0,
\end{align}
which resembles a linear equation $y=Vx+R_0$, where $R_0$ is the initial range (horizontal distance) between the vehicle and the light sensor. Hence, an improved speed estimation technique is to use linear regression by least-squares (LS) method which takes all the measurements into account and finds the linear trend by minimizing the residuals\footnote{Residual is the difference between an observed value and corresponding fitted value provided by a model.}. This method gives us both the speed of the vehicle ($V$) and the initial range ($R_0$).

For a curved road scenario, angle $\beta$ is defined as the angle that the end of the curved road and the vehicle headlight subtend at the center of the road curvature (see Fig.~\ref{fig:sysmodel_vehicle}) which changes with angular velocity ($\omega$) of the vehicle. After applying some geometry and trigonometry to the system model diagram for curved road, it can be shown that the channel model becomes
\begin{align}
   \label{eq:curvedeqn}
    P_r\left(\beta\right) = \frac{K\left(\cos \left(\beta/2\right)\right)^{n+1}}{\left(2r \sin \left(\beta/2\right)\right)^\gamma},
\end{align}
where $r$ is the known radius of curvature of the road~\cite{Abuella2019}. $R_{st}$ and $d_{st}$ in Fig.~\ref{fig:sysmodel_vehicle} were assumed to be negligible considering the car is at the beginning of the road curvature and the light sensor is close to the road. Next, the values of $\beta$ that minimize the squared errors between the simulated or measured received power and the received power from equation~(\ref{eq:curvedeqn}) for all measurements were estimated. After that, $\omega$ can be estimated from $\beta = \beta_0-\omega t$ using LS method as described before. Then, the linear speed ($V$) of the vehicle can be found from $V=\omega r$. The feasibility of both of these speed estimation techniques, for both straight and curved road scenarios, were proven through ray tracing simulation in~\cite{Abuella2019}.

\subsection{Gesture Recognition}
\label{subsec:gesture}
Non-contact hand gesture recognition has become an essential application of HCI, allowing computerized systems to perform complex tasks remotely by taking instructions through hand gestures. Gesture recognition technology can be broadly categorized into two types: wearable sensing and remote sensing. Wearable sensing provides better accuracy, but wearing sensors on the hands or fingers is inconvenient and can inhibit the spontaneous use of the technology. Remote or non-contact sensing, on the other hand, recognizes hand gestures without requiring any special hardware attached to the hands.

LWS has shown promise in non-contact gesture recognition, as demonstrated in~\cite{Yu2023} using visible light and ~\cite{Yu2021} using both visible light and infrared sensing. In these studies, human subjects performed eight common HCI hand gestures within 35 cm of the LWS setup, and the reflected light signals were captured as voltage difference waveforms. The stored waveforms were subjected to various signal processing techniques, including discrete wavelet denoising, time-domain thresholding, zero-padding, and Z-score standardization, to prepare the dataset for machine learning-based gesture classification. The data were then fed into a K-nearest neighbor (KNN) model~\cite{peterson2009} for classification, and model performance was evaluated using confusion matrices generated from 10-fold and leave-one-out cross validations. The best gesture recognition performance was achieved with infrared sensing at 20\,cm with
an average accuracy of 96\%~\cite{Yu2021}. Generally, infrared sensing outperformed visible light sensing in gesture recognition due to the peak responsivity of the light sensor being around the infrared wavelength used, resulting in higher received power and signal-to-noise ratio (SNR) with infrared light. The system model was evaluated under ambient lighting conditions, both on and off, in the room. Visible light sensing showed better gesture recognition accuracy in dark environments compared to bright environments. On the other hand, ambient lighting had little to no effect on the performance of infrared sensing, as the intensity of the reflected infrared light was much greater than the ambient contribution.

\subsection{Vitals Monitoring}
\label{subsec:vitals}
Assessing the physical and psychological states of a subject requires monitoring key vital signs, such as respiration rate and heart rate, and their associated patterns. Traditionally, these vital signs have been manually assessed by clinicians, which is neither accurate nor applicable to continuous monitoring. To address this, non-contact approaches for vitals monitoring have been developed that are preferred over contact-based or wearable approaches due to their convenience and wide-ranging applicability. LWS technology, which uses harmless and ubiquitous infrared or visible light, can detect variations in the reflected light signal due to human respiration and heartbeat from a distance. This technology is safer, simpler, faster, and more private than other technology counterparts like RF and optical camera-based vitals monitoring systems. Efforts have been made to successfully estimate respiration and heart rates and detect respiratory anomalies using LWS which are discussed next.

\subsubsection{\textbf{Respiration and Heart Rate Estimation}}
\label{subsubsec:resp_heart_rate}
Incoherent visible light sensing was utilized to estimate two important vitals parameters, respiration rate and heart rate, of human beings~\cite{Abuella2020, abuella2019conf}. In this study, human subjects were asked to breathe normally in seated positions in front of the VLS setup, while visible light was directed toward their chest. The received light intensity increases when the subject inhales and decreases when they exhale, resulting in data containing periodic respiration patterns modulated by heartbeats. The data were filtered using bandpass filters with appropriate cut-off frequencies and transformed into the frequency domain to find the frequency with the highest spectral amplitude ($f_{max}$), which denoted the respiration rate or heart rate. The estimated rates were compared to ground truths measured by U.S. Food and Drug Administration (FDA) approved contact-based devices. More than 94\% accuracy was achieved for both respiration and heart rate at 40\,cm test distance from the VLS setup.

Experiments were conducted with subjects in standing (applicable to the queue at airport security checkpoints) and supine positions (applicable to vitals monitoring during sleep or hospital setting) and less than 10\% error was noted in both cases. The system model was evaluated in ambient light on and off scenarios, with greater estimation accuracy achieved when ambient light was off due to less noise and interference from the dark environment. 

\subsubsection{\textbf{Respiratory Anomaly Detection}}
\label{subsubsec:resp_anomaly}
LWS can be used not only to estimate respiration and heart rate but also to detect anomalies in them. Infrared sensing can be preferred in human vitals monitoring since it is not visible to human eyes, hence more discreet and applicable to low-light environments. Respiratory anomaly detection was primarily targeted and its baseline was established by using a robot that could perform chest movements similar to humans during breathing, but with higher precision and repeatability~\cite{islam2023}. Breathing frequency and amplitude can be controlled precisely with the robot. Different arbitrary breathing patterns can be preloaded into the robot too, and its chest movement will follow that pattern. Besides normal breathing, known as Eupnea, 6 basic patterns of abnormal breathing (apnea, tachypnea, bradypnea, hyperpnea, hypopnea and Kussmaul’s breathing) were identified from the literature~\cite{Purnomo2021, Jagadev2020, Fekr2014, BarbosaPereira2017, Rehman2021, Parreira2010, Moraes2019}, and the robot was programmed to breathe in each of these patterns. One additional breathing class, faulty data, was included in the anomaly detection process so that the system could detect and discard the data that did not contain correct breathing information due to external or internal interruption/malfunction. Thus, breathing anomaly detection was reduced to an 8-class classification task.

Data pertaining to these 8 breathing classes were collected at three different test distances (0.5\,m, 1\,m and 1.5\,m) between the robot and the LWS setup. The data were pre-processed by filtering high-frequency noise, detrending to remove drift, and extracting handcrafted features for machine learning algorithms. Two ML algorithms, decision tree~\cite{song2015} and random forest~\cite{breiman2001}, were applied to the extracted features to classify the data, and model performance was assessed using average classification accuracies through 10-fold cross-validation. The decision tree model yielded the highest classification accuracy (96.6\%) with the data collected at 0.5\,m distance. Ensemble models such as random forest outperformed single models when evaluated with mixed data collected at different distances. The system model can be further improved by incorporating more variations in the training data and testing the system in a realistic environment using human subjects.

\begin{figure}[hbtp]
\includegraphics[width=.495\textwidth]{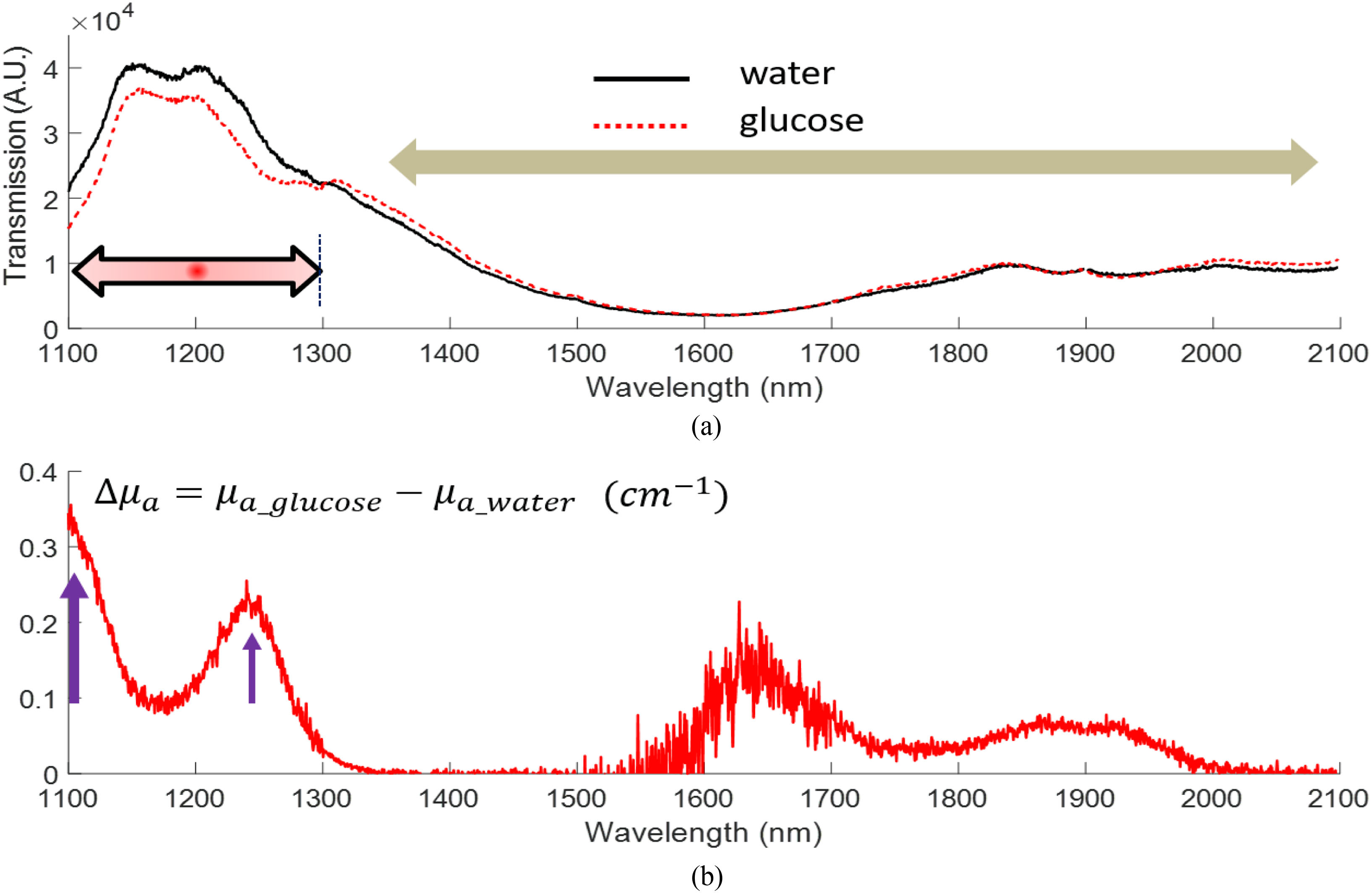}
\centering
    \caption{(a) The raw transmission spectra of the water and the glucose shot solution, (b) The differential spectral absorption of glucose that exceeds the water \cite{Piao2020}.}
\label{fig:glucose}
\end{figure}

\subsection{Glucose Sensing}
\label{subsec:glucose}
Measuring blood glucose levels is crucial in today's society due to the increasing prevalence of prediabetes and diabetes. Traditional methods of glucose measurement involve invasive blood collection, which can be inconvenient and uncomfortable for frequent use. To address this issue, researchers have been developing optical~\cite{optglu1, optnonoptglu1,optglu2,optglu3,optglu4,optglu5,optglu7} and non-optical~\cite{optnonoptglu1,nonoptglu2,nonoptglu3,nonoptglu4} non-invasive techniques for glucose sensing, where the probing devices come in contact with the skin. In contrast, incoherent light-based sensing has the potential to monitor blood glucose levels without physical contact, making it an ideal approach for discreet monitoring even during sleep. Its proof of concept has been demonstrated in~\cite{Piao2020} by measuring glucose concentration in a turbid aqueous medium from around 1\,m distance with a resolution of 8.9\,mg/dL. The interaction between human skin and light can be characterized by specular and diffuse reflections of light from the skin surface, light scattering from the surface and by wavelength-scale cellular bodies inside the tissue, and light absorption by chromophores that include water and glucose molecules. The key idea behind this technique was to identify a spectral window that significantly distinguishes light absorption by glucose and water for sensing purposes. In order to enhance glucose sensitivity, diffuse reflectance of light was utilized which allowed longer path-length of light inside tissue facilitating absorption by glucose. It has been shown in \cite{Piao2020} that the most favorable spectral window for non-contact glucose sensing has 1100-1300\,nm wavelengths, corresponding to the infrared light (Fig.~\ref{fig:glucose}). Therefore, incoherent infrared light-based sensing can be used for determining blood glucose levels without physical contact. 

\subsection{Occupancy Estimation}
\label{subsec:occupancy}
Indoor occupancy estimation and crowd counting have numerous applications, such as optimizing power consumption in buildings and designing marketing campaigns and emergency exit plans based on the spatial distribution of customers in retail stores or shopping malls. The number of people in an indoor area can be estimated by scanning information from Bluetooth low energy or radio frequency identification tags given to the occupants~\cite{basalamah2016,zhang2010}. But this approach has the overhead of distributing the tags to everyone and extra power consumption by the tags. Clearly, device-free methods for occupancy estimation are preferred, and LWS can play an important role. Unlike many other applications of LWS, deploying an extra light source is not always needed in this application because existing lights in the room can work as transmitters.

\begin{figure}[t]
\includegraphics[width=0.5\textwidth]{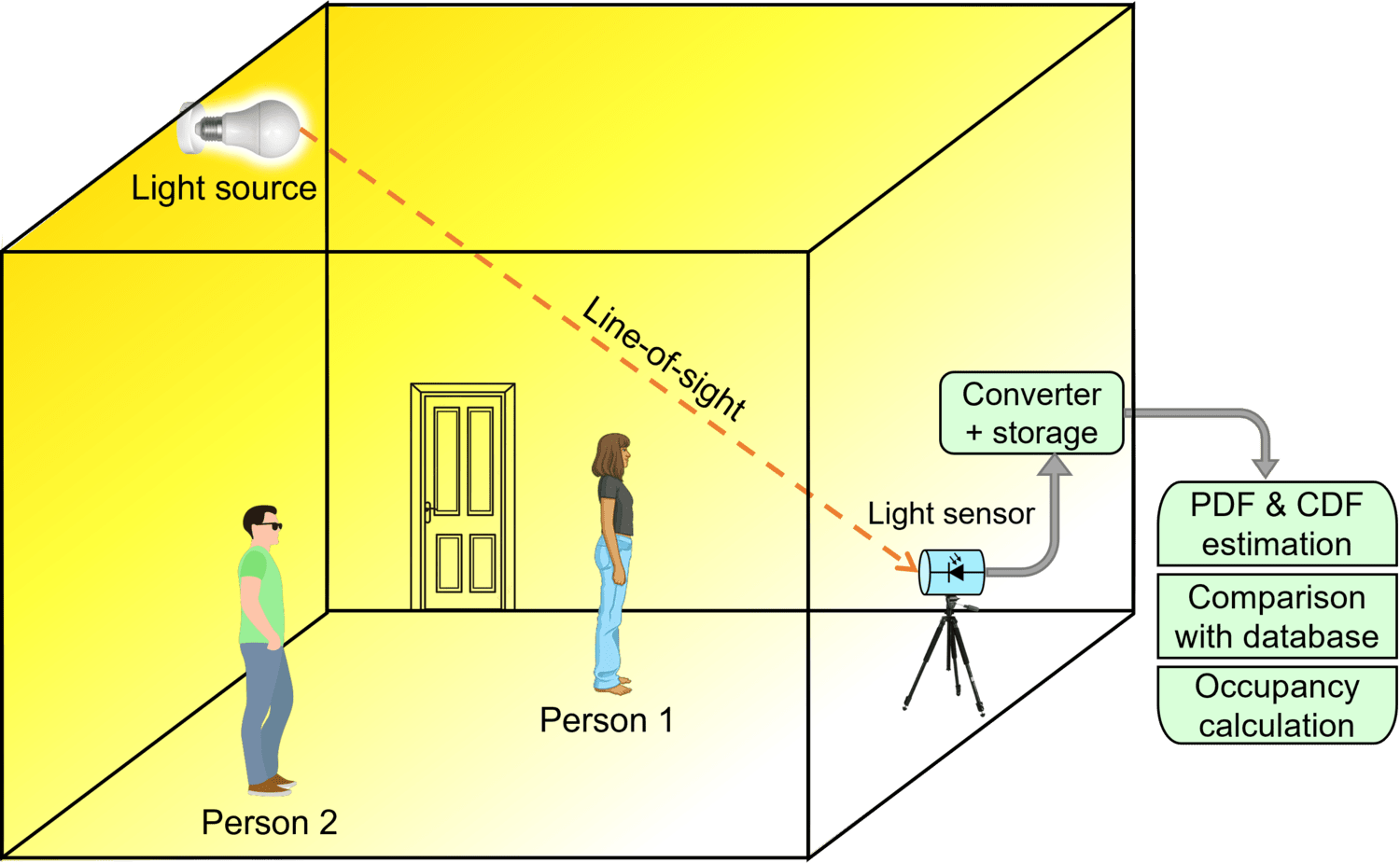}
\centering
    \caption{LWS system model for occupancy estimation.}
\label{fig:sysmodocc}
\end{figure}

Theoretical analysis and simulation of indoor occupancy estimation using visible light sensing have been performed in~\cite{Mestiraihi2018} using the LWS system model presented in Fig.~\ref{fig:sysmodocc}. The basic principle utilized in this work was that the probability density function (PDF) and cumulative distribution function (CDF)~\cite{miller2012} of the received light power change depending on the number of people in the room crossing the line-of-sight (LOS) between the light source and sensor. Analytical expressions of PDF and CDF of the received light power were derived which were functions of the probability of crossing the LOS by at least one person, the number of persons crossing the LOS, and a few other parameters. MATLAB simulation was performed by assuming reasonable values of the parameters to generate the same PDF and CDF. When the PDFs and CDFs were plotted, the theory and simulation were found to matched almost perfectly. To estimate occupancy, a PDF and CDF database were created for different known occupancy scenarios. For a new set of received power data generated through simulation, its PDF and CDF were compared with the labeled database using  Kullback-Leibler (KL) divergence method, and the occupancy was found from the number of people that generated the minimum divergence. However, this study did not consider the non-line-of-sight (NLOS) components of the received light power coming from scattering and multipath effects and left this, as well as experimental verification of the proposed method, as future work. But the agreement between theoretical and simulation results clearly proves the feasibility of applying LWS for occupancy estimation.

\subsection{Structural Health Monitoring}
\label{subsec:struct_health}
\begin{figure}[htbp]
\includegraphics[width=0.5\textwidth]{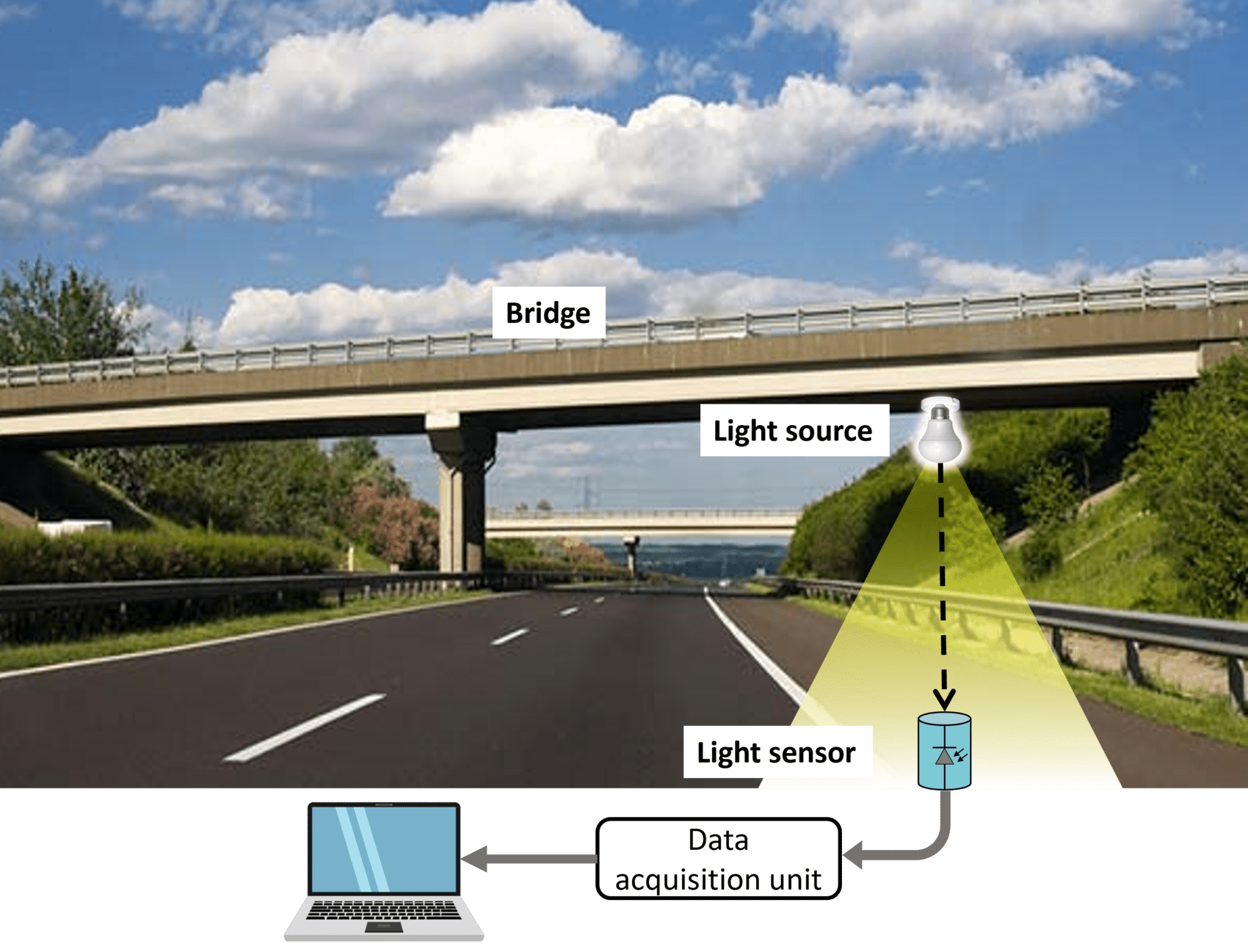}
\centering
    \caption{LWS system model for structural health monitoring of bridges (using direct light).}
\label{fig:sysmodstruct}
\end{figure}

Bridges are crucial components of transportation infrastructure and their full or partial failure can result in severe consequences for their users. Therefore, monitoring the deflections and vibration characteristics of bridges is crucial to ensure their safe operation and predict any damage that may occur. Various common technologies such as linear variable differential transformers (LVDT), GPS-based sensing, radar interferometry, laser vibrometers, and optical camera-based sensing have been used for structural health monitoring of bridges~\cite{joshi2017, yi2010, guan2014, miyashita2008, feng2015}. The LVDT installation and calibration are challenging because it requires a stationary point to attach the transducer with a fixed mounting structure. GPS-based monitoring suffers from low accuracy. The camera-based solution underperforms in inclement weather and low-lighting conditions. Radar and laser-based solutions have their drawbacks discussed in Section~\ref{sec:advantages_of_VILS} and Table~\ref{table:comparison}. Moreover, all of these solutions are expensive with the cost of required special equipment ranging from \$750 to \$3500. Visible light sensing with a simple LED and light sensor along with a data acquisition unit has been proposed to address the issues in the existing technologies in a cost-effective way ($<$\$400) while maintaining a sufficient level of accuracy in measuring bridge deflection~\cite{Soliman2019}. 
The LED can be attached to the measurement point of the bridge facing downward as shown in Fig.~\ref{fig:sysmodstruct} or placed on the ground facing upward with reflective material at the measurement point of the bridge. The light sensor detects the direct or reflected light intensity that contains the vibration characteristics of the bridge. The detected light intensity can be transformed to bridge displacement through calibration. Thus, the bridge displacement patterns can be stored and visualized for making decisions on the structural health of the bridge.

The proposed approach of structural health monitoring using VLS has been validated in~\cite{Soliman2019} through a laboratory setup consisting of two hydraulic actuators, two LVDTs and a transverse steel beam. The transverse beam could actuate the bridge displacement using a servo controller at various known maximum amplitudes and rates. The simple received power model $P_r=KD^{-\gamma}$ from equation~(\ref{eq:commonchannel}) has been utilized to quantify $K$ and $\gamma$ and calibrate the system using the collected data. The maximum displacement found using VLS and LVDT (the most accurate existing method) were compared and the error in VLS measurement was found to be $<$1.4\%.

\section{Limitations of Incoherent LWS}
\label{sec:limitations_of_VILS}
In addition to the benefits highlighted in Section~\ref{sec:advantages_of_VILS} and demonstrated in the applications examined in Section~\ref{sec:LWS_applications}, there exist certain constraints and implementation challenges of light-wave sensing and communication systems that have the potential to restrict their performance and effectiveness. Some of those limitations and corresponding implications are discussed as follows.

\begin{enumerate}[leftmargin=*]
\item \textbf{Limited modulation bandwidth}: Optical frequencies predominantly operate in the THz range, which offers high bandwidth for data transmission. However, the slow modulation response of LEDs due to their limited modulation bandwidth is a major concern~\cite{Karunatilaka2015,Abuella2021}. For instance, the modulation bandwidth of RGB LEDs is approximately 20\,MHz, which is inadequate for high-speed communication~\cite{zafar2017}.  However, this limitation may not be significant in wireless sensing applications that do not necessitate high-speed modulation of light. 

\item \textbf{Noise and interference from ambient light}: Despite optical frequencies being immune to interference with RF frequencies, ambient light and sunlight can still cause interference, leading to additive noise and a decrease in signal-to-noise ratio (SNR)~\cite{Tsiatmas2014, Karunatilaka2015, boucouvalas1996}. In particular, visible light is more susceptible to such interference during the daytime or when ambient light is present, as the contribution from the environment is primarily visible light. As discussed in the gesture recognition application in Section~\ref{subsec:gesture}, infrared sensing is less affected by ambient light due to the high responsivity of the light sensor in the IR range of frequencies and the narrower spectral width of IR compared to visible light.

\item \textbf{Shadowing}: Light-based sensing and communication systems are susceptible to the shadowing effect since light cannot pass through opaque objects~\cite{Abuella2021,xiang2014, Barry1995, Jovicic2013}. This can result in signal outages caused by objects or moving individuals in the surrounding area.

\item \textbf{Limited range}: Because of noise from ambient light and square-law nature of the light sensor that doubles the path loss (in dB) in the channel, VILS usually works only within a limited range~\cite{Barry1995}. In most of the healthcare applications of VILS reviewed in Section~\ref{sec:LWS_applications}, the sensing range was $\le$1\,m. One approach to extend the range is to modulate the transmitted light using a known carrier frequency and detect it using a coherent detection technique. Although the detection technique is called ``coherent", it does not imply that coherent light is being used, and thus all the benefits of incoherent light-wave sensing still remain applicable. It was applied in the respiratory anomaly detection work (discussed in Section~\ref{subsubsec:resp_anomaly}) using a lock-in amplifier~\cite{lockin1,lockin2} and the sensing worked from beyond 1.5\,m distance. Industry-grade lock-in amplifiers are generally expensive, but lock-in detection can be implemented in software or realized in analog circuitry, which can significantly reduce the cost.

\item \textbf{Non-linearity of LEDs}: The relationship between the current and the emitted light is non-linear for LEDs beyond a specific boundary~\cite{burchardt2014}. 
To prevent clipping or distortion in the transmitted and the received signals, intensity modulation must be performed within the LED's linear operational range~\cite{elgala2011,Karunatilaka2015}.
\end{enumerate}

\section{Conclusion}
\label{sec:conclusion}
Wireless sensing has emerged as a crucial technology for the betterment of human life in the contemporary world. Incoherent light has immense potential to meet the current need for faster and safer wireless sensing. The sensing applications discussed in this study demonstrate the potential to provide accurate, efficient, and user-friendly solutions that can outperform existing technologies such as RF, laser, or camera-based sensing. However, visible and infrared light sensing technology still faces various challenges that require further research and development to ensure its wider applicability.



\bibliographystyle{IEEEtran.bst}
\bibliography{references.bib}

\begin{IEEEbiography}[{ \includegraphics[width=1in,height=1.25in,clip]{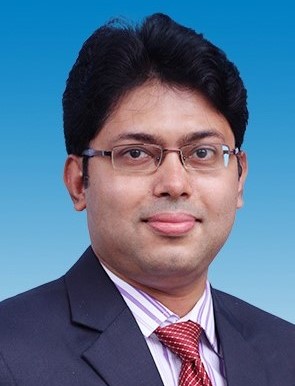}} ]{Md Zobaer Islam} received his B.Sc. degree in Electrical and Electronic Engineering in 2012 from Bangladesh University of Engineering and Technology, Dhaka, Bangladesh. He joined Oklahoma State University, Stillwater, OK as a graduate teaching and research assistant to pursue his Ph.D. degree at the School of Electrical and Computer Engineering in Spring 2020. He has industry experience of 4 years at Bangladesh Telecommunications Company Ltd. in the telecommunication and information technology (IT) sector and 3 years at Samsung R\&D Institute Bangladesh in the software sector. His current research interests include wireless light-wave sensing and machine learning.
\end{IEEEbiography}

\begin{IEEEbiography}[{\includegraphics[width=1in,height=1.25in,clip]{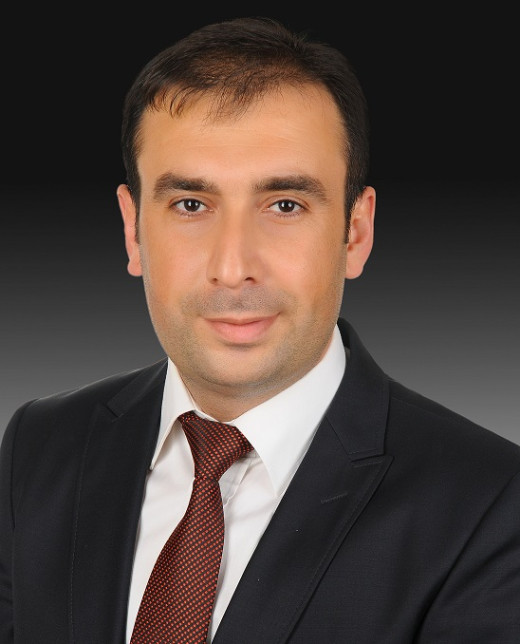}}]{Sabit Ekin (M'12, SM’21)} is currently working as an Associate Professor of the Department of Engineering Technology and Industrial Distribution at Texas A\&M University, College Station, Texas, USA. Previously, he was an associate professor of Electrical and Computer Engineering at Oklahoma State University (OSU). He  received the B.Sc. degree in electrical and electronics engineering from Eski\c sehir Osmangazi University, Turkey, in 2006, the M.Sc. degree in electrical engineering from New Mexico Tech, Socorro, NM, USA, in 2008, and the Ph.D. degree in electrical and computer engineering from Texas A\&M University, College Station, TX, USA, in 2012. He has four years of industrial experience as a Senior Modem Systems Engineer at Qualcomm Inc., where he has received numerous Qualstar awards for his achievements/contributions on cellular modem receiver design. His research interests include the design and analysis of wireless systems including mmWave and terahertz communications in both theoretical and practical points of views, visible light sensing, communications and applications, non-contact health monitoring, and Internet of Things applications.
\end{IEEEbiography}

\begin{IEEEbiography}[{ \includegraphics[width=1in,height=1.25in,clip]{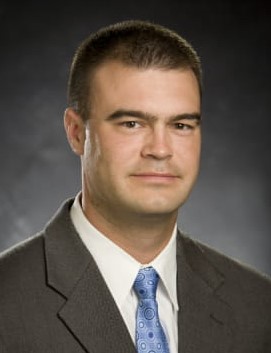}}]{John F. O'Hara (M'05, SM'19)} received his BSEE degree from the University of Michigan in 1998 and his Ph.D. (electrical engineering) from Oklahoma State University (OSU) in 2003.  He was a Director of Central Intelligence Postdoctoral Fellow at Los Alamos National Laboratory (LANL) until 2006.  From 2006-2011 he was with the Center for Integrated Nanotechnologies (LANL) and worked on numerous metamaterial projects involving dynamic control over chirality, resonance frequency, polarization, and modulation of terahertz waves.  In 2011, he founded a consulting/research company, Wavetech, LLC specializing in automation and IoT devices.  In 2017 he joined OSU Electrical \& Computer Engineering, where he is currently an assistant professor and the Jack H. Graham Endowed Fellow of Engineering.  His current research involves terahertz wireless communications, terahertz sensing and imaging with metamaterials, IoT, and light-based sensing and communications.  He has 4 patents and around 100 publications in journals and conference proceedings. 

\end{IEEEbiography}

\vfill

\end{document}